\documentclass[twocolumn]{aastex631} 

\newcommand{\hs}{h_{\rm s}}
\newcommand{\hc}{h_{\rm c}}
\newcommand{\ayr}{A_{\rm yr}}
\newcommand{\fyr}{f_{\rm yr}}

\newcommand{\mbulge}{M_{\rm bulge}}
\newcommand{\mbh}{M_{\bullet}}
\newcommand{\mmb}{\mbox{$M_{\bullet}$-$\mbulge$}}
\newcommand{\msig}{\mbox{$M_{\bullet}$-$\sigma$}}
\newcommand{\fbulge}{f_{\rm bulge}}
\newcommand{\fpair}{f_{\rm pair}}

\newcommand{\dist}{D_{\rm c}}

\newcommand{\fgearth}{f}
\newcommand{\fgrest}{f_{\rm r}}

\newcommand{\siginf}{\sigma_{\rm inf}}
\newcommand{\be}{\begin{equation}}
\newcommand{\ee}{\end{equation}}

\defcitealias{ss2016}{SB16}
\defcitealias{kh13}{KH13}
\defcitealias{mcconnellma13}{MM13}
\defcitealias{nml19}{dN19}

\begin{document}

\title{Exploring Proxies for the Supermassive Black Hole Mass Function: \\ Implications for Pulsar Timing Arrays}

\author[0000-0003-1407-6607]{Joseph Simon}
\altaffiliation{NSF Astronomy and Astrophysics Postdoctoral Fellow}
\affiliation{Department of Astrophysical and Planetary Sciences, University of Colorado, Boulder, CO 80309, USA}

\begin{abstract}

Supermassive black holes (SMBHs) reside at the center of every massive galaxy in the local Universe with masses that closely correlate with observations of their host galaxy implying a connected evolutionary history. The population of binary SMBHs, which form following galaxy mergers, is expected to produce a gravitational wave background (GWB) detectable by pulsar timing arrays (PTAs). 
PTAs are starting to see hints of what may be a GWB, and the amplitude of the emerging signal is towards the higher end of model predictions. Simulated populations of binary SMBHs can be constructed from observations of galaxies and are used to make predictions about the nature of the GWB. The greatest source of uncertainty in these observation-based models comes from the inference of the SMBH mass function, which is derived from observed host galaxy properties. 
In this paper, I undertake a new approach for inferring the SMBH mass function starting from a velocity dispersion function rather than a galaxy stellar mass function. I argue that this method allows for a more direct inference by relying on a larger suite of individual galaxy observations as well as relying on a more ``fundamental" SMBH mass relation. I find that the resulting binary SMBH population contains more massive systems at higher redshifts than previous models. 
Additionally, I explore the implications for the detection of individually resolvable sources in PTA data.

\end{abstract}

\section{Introduction}

Supermassive black holes (SMBHs), with masses of $10^6$ -- $10^{10} M_{\odot}$, reside in the nuclei of most, if not all, nearby galaxies. Shared evolution likely gives rise to the close correlation between the mass of the SMBH and several global properties of the host galaxy \citep{kormendy1995}. 
Following a galaxy merger, the SMBHs from each galaxy sink to the center of the common merger remnant through interactions with the galactic gas, stars and dark matter \citep{bbr80}. 
Once the SMBHs become gravitationally bound, they will emit strong gravitational waves (GWs) in the nanohertz frequency band. 
 
Utilizing a hierarchical framework of galaxy formation, many theoretical models of SMBH evolution have been used to infer the binary SMBH population and the resulting gravitational wave background (GWB) \citep[e.g.,][]{R+R95,jaffebacker,wyitheloeb,sesana+08,chen2019}. Full cosmological simulations have also been used to great effect from the Millennium simulation \citep{Sesana+09, Ravi+12} to the Illustris simulation \citep{kbh2017,skh20}. In recent years, focus has turned toward modeling the GWB based on the plethora of well-constrained galaxy observations \citep{sesana13,mcwilliams+14,ravi+15,mingarelli+17}. \citet[][hereafter \citetalias{ss2016}]{ss2016} undertook an in-depth analysis of how the uncertainty in galaxy observations propagated into the resulting GWB prediction and found that the inference of the SMBH mass function from host galaxy parameters had the largest impact. 

Over the past few decades, precision timing of millisecond pulsars has allowed the creation of a galactic-scale GW observatory, a pulsar timing array (PTA), which is sensitive to GWs at nanohertz frequencies \citep{NG12p5yr_data, IPTA_DR2_data}. The brightest signature detectable by PTAs is the GWB, which results from the incoherent superposition of GWs from SMBH binaries that form in post-merger galaxies. While the first stages of galaxy mergers have been repeatedly identified through electromagnetic observations \citep[e.g.,][]{dcm+2019, Stemo+20}, bound SMBH binaries, at sub-parsec separations, remain elusive. 
In fact, PTAs offer one of the most direct avenues to observing bound SMBH binaries \citep{AstroReview}. 

Recently, new PTA datasets are uncovering a signal which may be the first hints of the GWB \citep{ng12p5, PPTA_dr2_gwb, EPTA_dr2_gwb, IPTA_DR2}. If the signal is of an astrophysical origin, then confirmation of the signal's GW nature is expected in the next year or two \citep{astro4cast}. 
The emerging signal is towards the higher end of predicted amplitudes \citep{rosado+15,kbh2017,chen2019}, which may indicate that binary SMBHs are more massive than previously thought \citep{middleton+21} and/or that the local number density of SMBHs is larger than expected \citep{casey-clyde+21}. 
To help understand these implications, it is worth revisiting the inference of the SMBH mass function from galaxy observations. 

Inference of the SMBH mass function is predicated on measurements of host galaxy relationships in the local Universe (e.g., $\mmb$ and $\msig$). Previous work has used observations of the galaxy stellar mass function (GSMF) as the basis for the host galaxy relation used to infer the SMBH mass function \citep[][\citetalias{ss2016}]{sesana13}, however this method requires some assumption to be made about the fraction of the total stellar mass of the galaxy contained in the bulge, $\fbulge$. 
While the majority of contributions to the GWB is expected to be from bulge-dominated, early-type galaxies where $\fbulge \approx 1$ \citep{sesana13}, it is increasingly unclear how galaxy morphology tracks SMBH mass at higher redshift \citep{chen+20}, where galaxies tend to be smaller and clumpier \citep{conselice2014}.

A more direct inference could be made using a galaxy's stellar velocity dispersion ($\sigma$), which is measured from a galaxy's spectra. In addition to removing the need for an $\fbulge$ assumption, recent work suggests that $\msig$ is a more ``fundamental" relationship than $\mmb$ \citep[e.g.,][]{kh13, Bosch16, nml19, msg+20} and therefore more accurately predicts the SMBH mass. 
In particular the discovery of relic galaxies, such as NGC1271 and NGC1277 \citep{walsh2015,walsh2016}, which formed at $z > 2$ and have evolved passively ever since \citep{mateu2015}, points to the limitations of the $\mmb$ relation at higher redshifts. These relic galaxies are much smaller than most present day early-type galaxies, which is consistent with the strong redshift evolution of galaxy size \citep{vdWel2014}, however they host extremely massive SMBHs. While these galaxies are significant outliers on the $\mmb$ relation \citep{mateu2015}, they are consistent within the intrinsic scatter for the $\msig$ relation \citep{Bosch16}.

Until recently there were no spectroscopic surveys dedicated to measuring the galaxy velocity dispersion function (VDF) at higher redshifts, however, the Large Early Galaxy Astrophysics Census (LEGA-C) survey now provides the first direct spectroscopic measurement of the VDF from $0.6 < z < 1.0$ \citep{LEGAC2022}. The LEGA-C result is consistent with earlier attempts to indirectly measure the VDF using an \textit{inferred} velocity dispersion, calculated from photometric data \citep{Bezanson+11, Bezanson+12}, especially at large $\sigma$.
 
Previous attempts to use the $\msig$ relation in GWB calculations have started directly from the GSMF and used a galaxy stellar mass to $\sigma$ conversion derived from observations in the local Universe \citep{sesana13, sfb+16}, however, these results neglected to include the strong redshift evolution in the galaxy size--mass relation \citep{vdWel2014}. 
Thus, by incorporating more observational data, the VDFs used in this work provide fundamentally different starting points for calculating the SMBH mass function than anything that has been attempted before.

In this paper, I will use the VDFs from both \citet{Bezanson+12} and \citet{LEGAC2022} as the basis for inferring the SMBH mass function and compare the resulting binary SMBH population and GWB signal to that from the standard GSMF inference method. In the next section I describe my methods, including how to infer the SMBH mass function with these different approaches. I show my results in the following section, and close with a summary.

\section{Methods} \label{sec:methods}

Following \citetalias{ss2016}, this paper calculates the characteristic strain spectrum $\hc^{2} (f)$ from a cosmic population of binary SMBHs by integrating over all the sources emitting in an observed gravitational wave frequency bin $d\fgearth$ multiplied by the square strain of each source in that bin \citep[e.g.,][]{sesana+08},
\be
	\hc^{2} (\fgearth) = \fgearth \int \int \int dz ~dM_{\bullet} ~dq_{\bullet} ~\hs^{2} ~\frac{d^{4}N}{dz ~dM_{\bullet} ~dq_{\bullet} ~d\fgearth},
	\label{eqn:CharStrain_d4N}
\ee
where $d^{4}N$ is explicitly the number of binaries in a given redshift range $dz$, primary black hole mass range $dM_{\bullet}$, and black hole mass ratio range $dq_{\bullet}$, which are emitting in a given Earth-observed GW frequency range $df$, and $\hs$ is the polarization- and sky-averaged strain contribution from each binary \citep[e.g.,][]{thorne87},
\be
    \hs = \sqrt{\frac{32}{5}} \left(\frac{G M_{c}}{c^{3}} \right)^{5/3} \frac{\left( \pi \fgrest \right)^{2/3} c}{\dist}~,
    \label{eqn:strain}
\ee
where $M_c = \mbh (q_{\bullet}^{3}/(1+q_{\bullet}))^{1/5}$ is the chirp mass of the binary, $\dist$ is the proper (co-moving) distance to the binary, and $\fgrest$ is the frequency of the GWs emitted in the rest frame of the binary. The Earth-observed GW frequency $\fgearth = \fgrest / (1+z)$.

The number of binaries $d^{4}N$ is directly related to the comoving number density per unit redshift, primary black hole mass and binary mass ratio, 
\be
    \frac{d^{4} N}{dz ~dM_{\bullet} ~dq_{\bullet} ~d\fgearth} = \frac{d^{3} n}{dz ~dM_{\bullet} ~dq_{\bullet}} ~\frac{dV_{c}}{dz} ~\frac{dz}{dt} ~\frac{dt}{d\fgearth}.
	\label{eqn:d4N_BH}
\ee
The conversion above takes the number of binaries per co-moving volume shell, $dV_{c}$, and converts it to the number of binaries per Earth-observed GW frequency bin, $d\fgearth$, by first converting to redshift, and then to Earth-observed time. Once the binary hardens and decouples from the surrounding galactic environment, the orbital changes of the system become dominated by the emission of gravitational radiation at a rate of \citep{pm63},
\be
    \frac{df_{\rm orb}}{dt} = \frac{96}{5} \left( \frac{G M_{c}}{c^{3}} \right)^{5/3} ( 2 \pi )^{8/3} f_{\rm orb}^{11/3}.
    \label{eqn:dfdt}
\ee
For binaries in a circular orbit, the frequency of GWs emitted in the rest frame of the binary $\fgrest = 2 f_{\rm orb}$.

The frequency dependence of $\hc$ is captured in both $\hs$ and $dt/df$, therefore by combining Eqns. \ref{eqn:CharStrain_d4N}-\ref{eqn:dfdt}, $\hc(f)$ can be written as a simple power-law with dimensionless amplitude $\ayr$ referenced to the frequency of an inverse year $\fyr = 1$ yr$^{-1}$, \citep{jenet+06}.
\be
    h_{c} (f) = \ayr \left( \frac{\fgearth}{\fyr} \right)^{-2/3}.
\ee
However, this only applies for circular binaries in GW-radiation driven inspirals. If the inspirals were instead driven by some environmental coupling in the lowest frequencies of the PTA band, this would cause a turn-over in the power-law spectral shape \citep{sampson+15}. 
Environmental coupling can also impact the eccentricity of the binary and therefore the spectral shape of the characteristic strain spectrum in the PTA band \citep{ravi+14, huerta+15}. For the purposes of this project, where I am solely focused on the SMBH mass function, I ignore the specifics of environmental coupling as well as eccentricity and assume that all binaries remain in circular orbits while emitting GWs in the PTA band. 

\subsection{Determining the Number Density of Binary SMBHs} \label{sec:smbh_infer}

The number density of binary SMBHs is a combination of two functions: the SMBH mass function $\Phi_{\bullet} (z, M_{\bullet})$ and the binary SMBH merger rate $\mathcal{R}_{\bullet} (z, M_{\bullet}, q_{\bullet})$,
\be
	\frac{d^{3}n_{\bullet}}{dz ~dM_{\bullet} ~dq_{\bullet}} = \Phi_{\bullet} (z, M_{\bullet}) ~\mathcal{R}_{\bullet} (z, M_{\bullet}, q_{\bullet}).
	\label{eqn:d3n_BH}
\ee
There are currently no direct observational constraints on the demographics of binary SMBHs, so these functions are not explicitly known. Instead, they need to be inferred from other data. 
As in \citetalias{ss2016}, I use the demographics of galaxy mergers as a proxy. 

The binary SMBH merger rate is directly related to the galaxy merger rate, $\mathcal{R}_{\bullet} \sim \mathcal{R}_{\rm Gal}$, but offset in redshift due to the binary evolution timescale. 
The standard way to calculate the galaxy merger rate, $\mathcal{R}_{\rm Gal}$, directly from observations is by combining a galaxy close pair fraction, $\fpair$, with a merger timescale, $\tau$. The galaxy close pair fraction is an astronomical observable, while the merger timescale, which approximates the amount of time galaxy close pairs are observable, must be determined through simulations \citepalias{ss2016}.

The SMBH mass function, $\Phi_{\bullet}$, is more difficult to derive, and needs to be indirectly inferred by populating each galaxy with a central SMBH assuming some relationship between SMBHs and their host galaxies (e.g., $\msig$ relation, $\mmb$ relation, etc.). 
The standard approach to inferring the SMBH mass function from galaxy surveys is to start from an observed galaxy stellar mass function (GSMF), then make some assumption about the fraction of the total stellar mass of the galaxy contained in the bulge, $\fbulge$, and finally calculate the SMBH mass from the $\mmb$ relation \citep[][\citetalias{ss2016}]{sesana13, ravi+15}. 
Using this approach, the number density of binary SMBHs is directly inferred from the number density of galaxy mergers,
\be
   	\frac{d^{3}n_{\bullet}}{dz ~dM_{\bullet} ~dq_{\bullet}} = \left| \Phi_{\rm Gal} ~\mathcal{R}_{\rm Gal} \right|_{\mathrm{Gal} \rightarrow \bullet}.
	\label{eqn:d3n_Gal} 
\ee
where $(\mathrm{Gal} \rightarrow \bullet)$ represents the step where the SMBH is populated into the galaxy assuming some galaxy number density parametrization $\Phi_{\rm Gal}$.
When using a GSMF, $\Phi_{\rm Gal} = \Phi_{\rm GSMF} (z, M)$, where $M$ is the total stellar mass of the primary galaxy. The companion galaxy then has a mass of $q_{M} M$, where $0.25 < q_{M} < 1$ is set to capture major merger events, and the $(\mathrm{Gal} \rightarrow \bullet)$ step in Equation \ref{eqn:d3n_Gal} utilizes the $\mmb$ relation and includes the determination of $\fbulge$ for each galaxy. 

While the specific prescription for $\fbulge$ can change the predicted amplitude of the GWB by a factor of two or more, the largest source of uncertainty in this inference of the SMBH mass function inference comes from uncertainty in the GSMF itself \citepalias{ss2016}. This is to be expected given discrepancies over the number density of the most massive galaxies in the local Universe \citep{bernardi+13}. 
In addition to the issues with properly identifying massive galaxies, there is increasing evidence that SMBH evolution and bulge evolution track differently at higher redshifts ($z > 1$) \citep{chen+20}, however, there is no significant evidence that these different evolutionary tracks are caused by a change in the relationship between SMBH mass and other host galaxy properties \citep{sgh+15, sct+20}. 
Ideally, one would like a more robust method for inferring the SMBH mass function, particularly something that doesn't require an $\fbulge$ determination and doesn't have as much dependency on the underlying mass-to-light (M/L) model used to determine stellar mass \citep{bernardi+13}. 

The stellar velocity dispersion ($\sigma$) appears to allow for a more direct inference of SMBH mass. Recently, it has been suggested that $\sigma$ is a more fundamental property of a galaxy than it's stellar mass \citep{Bosch16, nml19, msg+20}, and that the $\mmb$ relation is just an extension of the more fundamental $\msig$ relation \citep{wdf12}.
However, it has been difficult to obtain VDFs, specifically at higher redshifts ($z > 0.3$), due to a lack of complete spectroscopic surveys to use as a separate starting point for inferring the SMBH mass function. Instead, when $\sigma$ has been explored in the context of simulating the binary SMBH population, it has been inferred from galaxy total stellar mass, and therefore, the results have unsurprisingly been consistent \citep{sesana13, sfb+16}. 
Instead, in an effort to reduce uncertainty in the inferred SMBH mass function, we are pursuing a novel approach to modeling $M_{\bullet}$ using the velocity dispersion functions from both \citet{Bezanson+12}, which are derived from an \textit{inferred} velocity dispersion ($\siginf$), and \citet{LEGAC2022}, which are derived from spectroscopic observations at $0.6 < z < 1.0$.

\subsection{The Velocity Dispersion Function}\label{sec:vdf}

When using a VDF in Equation \ref{eqn:d3n_Gal}, $\Phi_{\rm Gal} = \Phi_{\rm VDF} (z, \sigma)$, where $\sigma$ is the velocity dispersion of the primary galaxy. The companion galaxy then has a velocity dispersion of $q_{\sigma} \sigma$, where $q_{\sigma}$ is set to capture major mergers the same way that $q_{M}$ was in the case of GSMFs. Finally, the $(\mathrm{Gal} \rightarrow \bullet)$ step in Equation \ref{eqn:d3n_Gal} now utilizes the $\msig$ relation and no longer requires any additional steps as it did in the case of GSMFs.
\citet{LEGAC2022} measures the VDF directly through spectroscopic observations. \citet{Bezanson+11} uses a suite of photometric data to calculate an \textit{inferred} VDF, which is shown to be comparable with a spectroscopic measurements in the local Universe.

In \citet{Bezanson+11}, the \textit{inferred} velocity dispersion, $\siginf$, is measured from a combination of a galaxy's photometric properties. Starting from the virial theorem and building from the observational evidence that for local galaxies stellar mass is proportional to dynamical mass \citep{taylor+10}, the central velocity dispersion, $\sigma$, of a galaxy can be effectively predicted based on its size (effective radius, $r_{e}$), shape (S\`ersic indec, $n$), and total stellar mass ($M_{\ast}$).
\be
    \sigma_{\rm inf} = \sqrt{ \frac{G ~ M_{\ast}} {K_{\ast}(n) ~r_{e}} },
    \label{eq:sigma_inf}
\ee
where $K_{\ast}(n) \equiv K_{v} (n) \left( M_{\ast} / M_{\rm dynamical} \right)$. $K_{v} (n)$ is a S\`ersic dependent virial constant \citep{bcd02}, and since \citet{taylor+10} has shown that $K_{\ast}(n)$ only weakly depends on mass, the average ratio of stellar to total mass is adopted and calibrated such that the median $\siginf$ equals the median measured $\sigma$ in SDSS data. 

\citet{Bezanson+11} finds a scatter in the $\siginf$ - $\sigma$ relationship of $0.06$ dex, which has the effect of increasing the high-$\sigma$ end of the $\siginf$ distribution. This scatter must be properly taken into account when inferring the ``true" velocity dispersion function (VDF) in order to not over predict the number density of massive galaxies. Numerical fits for the ``true" VDF were not published in either \citet{Bezanson+11} or \citet{Bezanson+12}, and instead were obtained through private communication with the author \citep{Bezanson_Email}. The fits are done using a modified Schechter function and, given the strong correlation between $\sigma_{\ast}$, $\alpha$, and $\beta$ \citep{sheth+03}, uncertainties on the fit were obtained by setting $\alpha$ and $\beta$ to their maximum likelihood values and allowing $\sigma_{\ast}$ to absorb the combined uncertainties in those three parameters.

\subsection{Observational Constraints} \label{sec:obs}

Unlike previous work \citepalias[e.g.,][]{ss2016}, this paper is focused on keeping as many things constant in the calculation of the GWB as possible, in order to highlight the differences in the SMBH mass function and the implications for PTA experiments. As such, this work only uses the galaxy close pair fraction from \citet{keenan+14} in it's calculation of $\mathcal{R}_{\rm Gal}$, as opposed to \citet{robotham+14} since \citetalias{ss2016} showed them to be consistent. For $\tau$, the same formulation as \citetalias{ss2016} is used, which combines the mass and redshift dependence from \citet{kw08} with the results from \citet{lotz+11}.

\citet{Bezanson+12} calculates the VDF at $0.3 < z < 1.5$ using photometric measurements of galaxy stellar mass, effective radii, and S\`ersic index for the galaxies in the NMBS COSMOS \citep{NMBS_COSMOS} and the UKIDSS UDS fields \citep{UKIDSS}. In an attempt to make sure that the galaxy samples are as consistent as possible, this paper compares the VDF derived from the above $\siginf$ data to the GSMF from \citet{tomczak+14} since it is partially based on NMBS COSMOS. It is important to also note that the stellar mass determination in \citet{Bezanson+12} and \citet{tomczak+14} use the same software pipeline, which should reduce the amount of bias inserted into these calculations from different M/L determinations. 
When using the spectroscopic VDF from \citet{LEGAC2022} for $0.6 < z < 1.0$, the \textit{inferred} VDF is used at $z > 1$ since the \textit{inferred} VDF matches the spectroscopic VDF.

For inference of the SMBH mass function at $z < 0.3$, this paper uses the GSMF in \citet{moustakas+13} calculated from SDSS data since that is what was used in \citetalias{ss2016}, and it was also used as the local Universe comparison in \citet{tomczak+14}. To minimize introducing additional bias, the VDF from \citet{bernardi+10} is used at $z < 0.3$, since it was directly derived from similar SDSS data as \citet{moustakas+13}. 

To infer the SMBH in each galaxy, the $\mmb$ relation is used with the GSMF while the $\msig$ relation is used with the VDF.
The one place this paper deviates from holding many things constant is when determining which $\mmb$ and $\msig$ relations to use, since this parameter has been shown to have the largest impact on the predicted amplitude of the GWB \citepalias{ss2016}. Thus, relations are used from \citet[][hereafter \citetalias{mcconnellma13}]{mcconnellma13}, \citet[][hereafter \citetalias{kh13}]{kh13}, and \citet[][hereafter \citetalias{nml19}]{nml19} as these papers are representative of the range of values these relations have been shown to have.
It is worth noting that the range of measurements for the $\msig$ relation is smaller than that for the $\mmb$ relation adding further evidence for the $\msig$ relation to be considered more ``fundamental'' than the $\mmb$.

\subsection{Calculating the GWB}

As in \citetalias{ss2016}, Eqs. \ref{eqn:CharStrain_d4N}, \ref{eqn:d4N_BH}, and \ref{eqn:d3n_Gal} are combined to calculate $\hc$. When using a GSMF,
\begin{equation}
    \hc^{2} = f \int \int \int ~\frac{\Phi_{\rm GSMF}}{\tau} \frac{df_{\rm pair}}{dq_{M}} \frac{dV_{c}}{dz} \left( \frac{dt}{df} h_{s}^{2} \right) dz~dM~dq_{M}.
    \label{eqn:hc_gsmf}
\end{equation}
When using the VDF, as described in \S \ref{sec:vdf},
\begin{equation}
    \hc^{2} = f \int \int \int ~\frac{\Phi_{\rm VDF}}{\tau} \frac{df_{\rm pair}}{dq_{\sigma}} \frac{dV_{c}}{dz} \left( \frac{dt}{df} h_{s}^{2} \right) dz~d\sigma~dq_{\sigma}.
    \label{eqn:hc_vdf}
\end{equation}
To calculate $\ayr^2$, the above integrals are used setting $f = \fyr$. 

The limits of integration for the above equations are chosen in order to capture the major galaxy mergers in the local Universe, which make up the majority of contributing systems to the GWB in the PTA band. For the GSMF (Eqn \ref{eqn:hc_gsmf}) in \citetalias{ss2016}, these are shown to be $0 < z < 1.5$, $10 < \mathrm{log}_{10} M_{\rm Gal} < 12.5$, and $0.25 < q < 1$. When using the VDF, it is imperative to ensure that the integration happens over a simiilar range of galaxy mergers. To do so, this work uses the size - mass relation from \citet{williams2010} because it is based on the same UKIDSS sample that the VDF in \citet{Bezanson+12} is derived from. The corresponding limits of integration for the VDF are then $0 < z < 1.5$, $1.85 < \mathrm{log}_{10} \sigma < 2.6$, $0.67 < q_{\sigma} < 1$. While these ranges are smaller, it is important to remember that the corresponding range of $\sigma$ values is much narrower for a given range of $M$. Integration limits derived from the size-mass relation in \citet{shen+03}, which is based on SDSS data, were also tested, but produced almost no change in the VDF predictions for $\hc$.

\begin{figure}
    \begin{center}
    \includegraphics[width=\columnwidth]{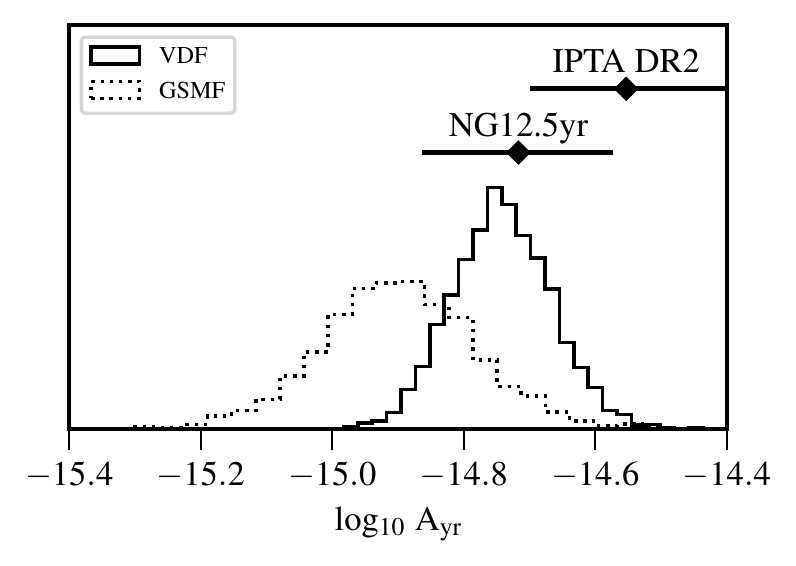}
    \end{center}
    \caption{$\ayr$ predictions calculated using the VDF (solid) and GSMF (dotted) methods to infer the SMBH mass function. The constraints from both the NANOGrav 12.5-year dataset \citep{ng12p5} and IPTA DR2 \citep{IPTA_DR2} are shown above the two histograms.
    Each histogram includes $3000$ predictions, which $1000$ coming from each of the three SMBH -- host galaxy relations used. The $90\%$ interval of the VDF predictions cover $0.24$ dex with a mean at log$_{10} \,\ayr = -14.74$, while the GSMF predictions cover $0.4$ dex over the $90\%$ interval with a mean at log$_{10} \,\ayr = -14.9$. Both methods are able to reproduce the emerging signal, but the VDF predictions tend towards higher amplitudes than the GSMF predictions.
    }
    \label{fig:AyrCompare}
\end{figure}

\begin{figure*}
    \centering
    \gridline{
        \fig{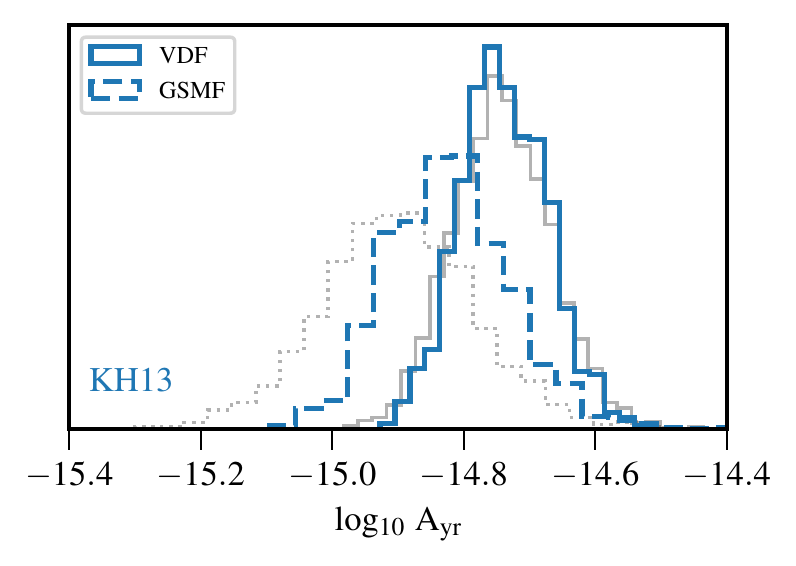}{0.32\textwidth}{}
        \fig{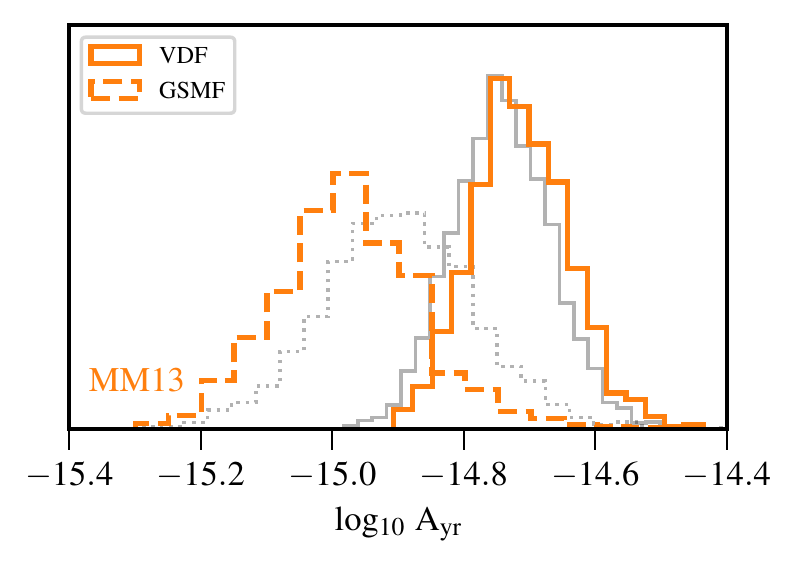}{0.32\textwidth}{}
        \fig{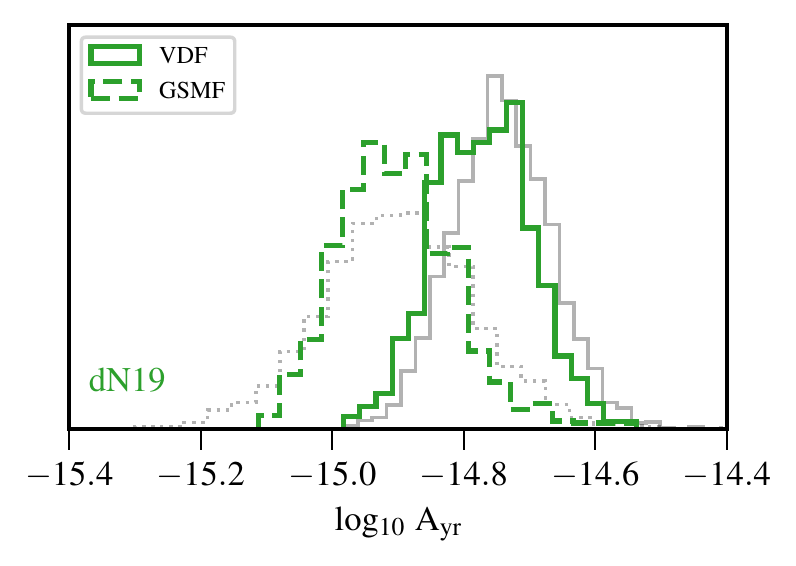}{0.32\textwidth}{}}
    \caption{
    $\ayr$ predictions calculated using the VDF (solid) and GSMF (dashed) methods to infer the SMBH mass function. Each panel shows the predictions for a different SMBH -- host galaxy relation: \citetalias{kh13} is shown in blue in the left panel, \citetalias{mcconnellma13} is shown in orange in the center panel, and \citetalias{nml19} is shown in green in the right panel. Additionally, in all three panels, the full distribution of predictions from \autoref{fig:AyrCompare} for all three relations is shown in gray. The VDF predictions appear to not be sensitive to the choice of host galaxy relation in the same way that the GSMF predictions are, and as seen in \autoref{fig:AyrCompare} the VDF predictions are always at a higher amplitude. However, the magnitude of the difference between the VDF and GSMF predictions is dependent on the specific host galaxy relation chosen. The largest difference is seen when using \citetalias{mcconnellma13} and the smallest difference is seen when using \citetalias{kh13}.
    }
    \label{fig:AyrBreakdown}
\end{figure*}

\begin{figure*}
  \plottwo{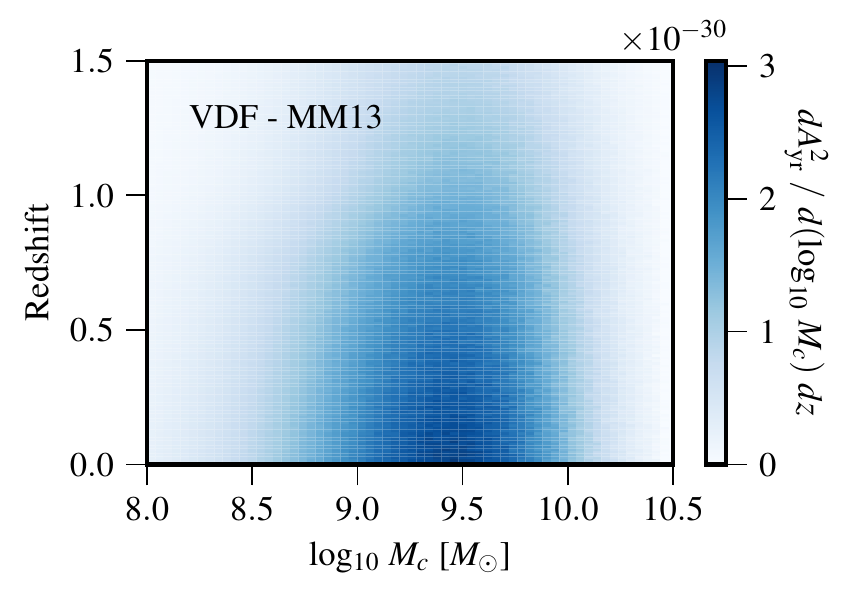}{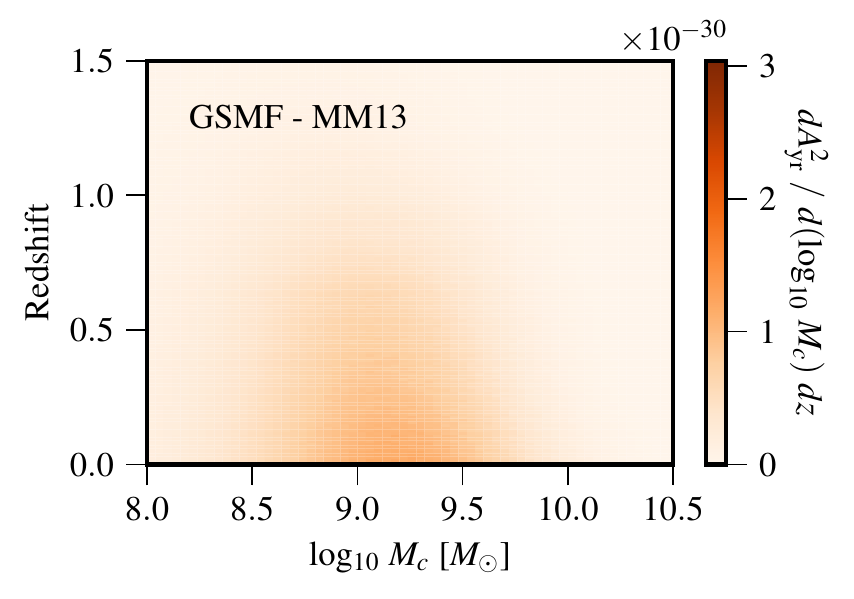}
  \plottwo{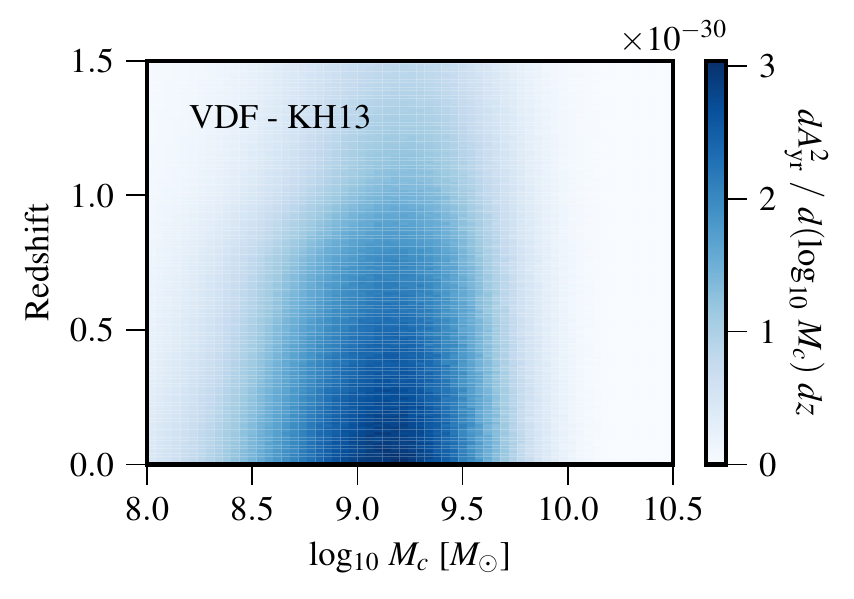}{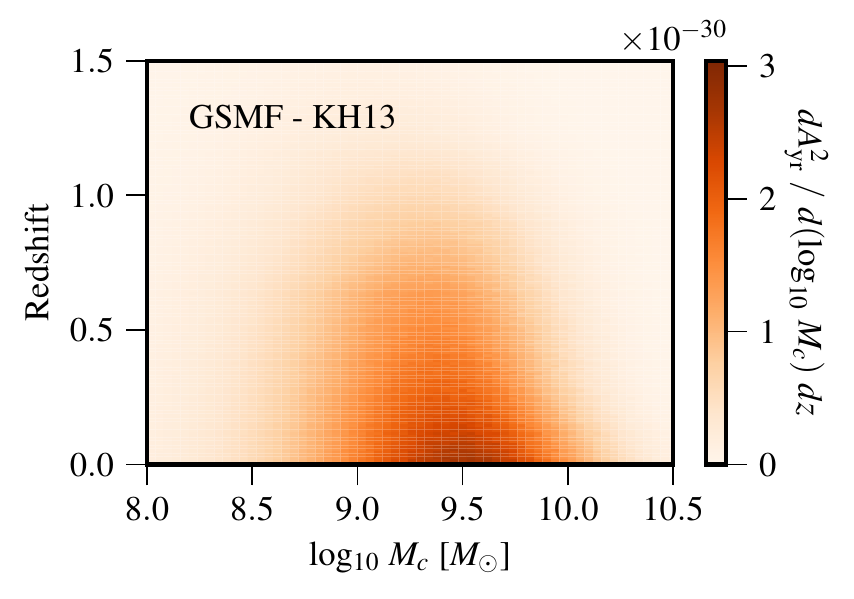}
  \caption{
  2D differential contribution to $\ayr^2$ from two different aspects of the simulated binary SMBH population: binary chirp mass ($M_{c}$) shown on the x-axis and redshift shown on the y-axis. The columns show the VDF inference on the left (in blue) and the GSMF inference on the right (in orange). The rows represent two different SMBH -- host galaxy relations with \citetalias{mcconnellma13} on the top and \citetalias{kh13} on the bottom. The limits of the colorbars are the same in all cases. The biggest difference between the VDF and GSMF is the contribution to the signal at higher redshift ($z > 0.5$). In the case of the GSMF inference, the population of binaries that contribute significantly to the GWB is contained to the local Universe, while the VDF inference shows significant contributions all the way out to $z \sim 1$ and beyond. One can infer from this that the underlying SMBH mass function implied by the VDF includes more massive systems at higher redshifts then that implied by the GSMF. 
  }
  \label{fig:Ayr2Dcompare}
\end{figure*}

\section{Results}

Combining all of the observational constraints from \S\ref{sec:obs} into the models in \autoref{eqn:hc_gsmf} and \autoref{eqn:hc_vdf}, I calculate predictions for the GWB. For $\ayr$ predictions, I generate 1000 realizations each for the VDF and GSMF under each SMBH -- host galaxy relation. In each realization, all parameters are allowed to vary within their observed uncertainties.

\pagebreak
\subsection{Predicted Amplitude of the GWB}

\autoref{fig:AyrCompare} shows the relative distributions predicted for $\ayr$ from the two different methods (GSMF and VDF) employed in this paper. The predictions based on the VDF are both at a higher amplitude than those from the GSMF and cover a smaller spread implying that the underlying binary SMBH populations inferred by these two methods are quite different. As discussed in \S\ref{sec:obs}, I have endeavored to minimize any differences in observational inputs for these predictions, in order to highlight that the subsequent $\ayr$ differences are due primarily to differences in the method used to infer the SMBH mass function (discussed in \S\ref{sec:smbh_infer}), and not simply due to using different observational inputs to the model. 
The larger values of $\ayr$ predicted by the VDF are more in line with the amplitude of the common process starting to appear in PTA data, however, both methods are able to reproduce the signal emerging in PTA datasets.
To highlight this, \autoref{fig:AyrCompare} also includes the $\ayr$ constraints from both the NANOGrav 12.5yr dataset \citep{ng12p5}, as well as IPTA DR 2 \citep{IPTA_DR2}. Yet, without the requisite spatial correlations the origin of the signal appearing in PTA data remains unclear.

Interestingly, the VDF predictions for $\ayr$ appear to be less sensitive to the choice of SMBH -- host galaxy relation than the GSMF method. In \autoref{fig:AyrBreakdown}, the three panels show the $\ayr$ predictions for each SMBH -- host galaxy relation used in this work. In all three cases, the VDF predictions are nearly identical and almost perfectly match the combined distribution seen in \autoref{fig:AyrCompare}. In contrast, the mean value of the GSMF prediction changes significantly depending on which SMBH -- host galaxy relation is utilized. This behavior of the GSMF predictions has been previously noted in \citetalias{ss2016}.

One of my aims for pursuing this new method of inferring the SMBH mass function was to reduce the uncertainty on $\ayr$ that is present under the GSMF method. Some of the reduction in uncertainty likely comes from removing the intermediate step of calculating $\fbulge$ for each galaxy. $\fbulge$ covered a broad range of values due to limited observational evidence, however, most systems that strongly contribute to the GWB likely have $\fbulge \sim 1$ \citep{sesana13, kbh2017}.
Some of the reduction in uncertainty is due to the narrower spread of measurements for the $\msig$ relation in comparison to the $\mmb$ relation. 
Either way, it appears that by including observations of both a galaxy's size and S\'ersic index along with stellar mass in the inference of the central SMBH, the VDF provides a more direct avenue to the SMBH mass function. 

To help understand where the differences in the $\ayr$ predictions are coming from, \autoref{fig:Ayr2Dcompare} shows the 2D differential contribution to $\ayr^2$ from $M_{c}$ and $z$.
The most striking difference is the contribution to $\ayr^2$ that the VDF receives at higher redshifts ($z > 0.5$). 
\autoref{fig:Ayr2Dcompare} also shows that the additional contribution at higher redshift is present regardless of which SMBH -- host galaxy relation is used with the VDF method. Additionally, one sees that the different SMBH -- host galaxy relations impact the predicted SMBH masses from each method in opposite ways. For instance, the relations in \citetalias{mcconnellma13} predict higher SMBH masses when using $\msig$ than when using $\mbulge$, and that is reversed when the relations in \citetalias{kh13} are used, with the GSMF method producing more massive SMBH systems, especially in the local Universe ($z < 0.2$). It is obvious just from looking at the different scaling in the colormaps between the two GSMF models that the predicted amplitude using \citetalias{mcconnellma13} is lower than \citetalias{kh13}. This is in contrast to the similarities in the amplitude predictions from the VDF models, which can be seen in \autoref{fig:AyrBreakdown}.
Overall, one can infer that the SMBH mass function implied by the VDF includes more massive SMBHs at higher redshifts than the mass function implied by the GSMF. 

The results from using the VDF model are similar to those found in \citet{casey-clyde+21}, which uses a quasar-based model to predict the binary SMBH population. In that case, the SMBH mass function has more support at larger mass values than the GSMF models from \citet{sesana13} that they compare to, the amplitude of the GWB is more in line with the NANOGrav 12.5yr dataset, and the redshift contribution from the quasar-based model is more evenly distributed like the VDF model shown here. 

\begin{figure}
    \begin{center}
    \includegraphics[width=\columnwidth]{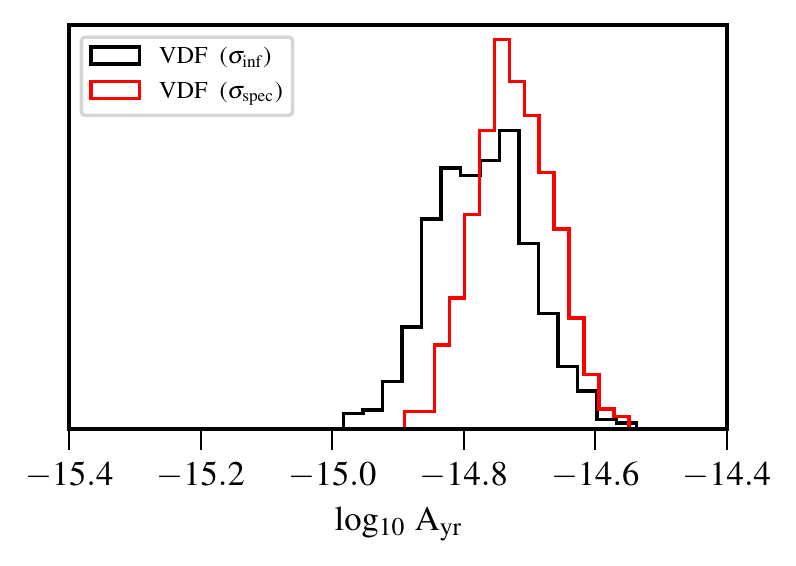}
    \end{center}
    \caption{
    $\ayr$ predictions calculated using the \textit{inferred} VDF ($\sigma_\mathrm{inf}$, black line) and the spectroscopic VDF ($\sigma_\mathrm{spec}$, red line) to infer the SMBH mass function. In the case of the $\sigma_\mathrm{spec}$ predictions, spectroscopic data from SDSS \citep{bernardi+10} and the LEGA-C survey \citep{LEGAC2022} were used at $z < 1$, while the \textit{inferred} VDF was used to fill in from $1 < z < 1.5$. Given the consistency that the LEGA-C survey finds between the spectroscopic and \textit{inferred} VDFs \citep{LEGAC2022}, it is not surprising that these results are so consistent. While it would be preferred to use only spectroscopic VDFs for this work, the similarity in these results show that the \textit{inferred} VDFs are an acceptable alternative.
    }
    \label{fig:AyrVDFCompare}
\end{figure}

\subsection{Spectroscopic vs. Inferred VDF}

As discussed in \S\ref{sec:vdf}, this paper primarily uses the \textit{inferred} VDF from \citet{Bezanson+12}, however, given the recent publication of the spectroscopic VDF observed with the LEGA-C survey \citep{LEGAC2022}, I calculate $\ayr$ using both functions. \autoref{fig:AyrVDFCompare} shows the relative distributions predicted for $\ayr$ under each model, however unlike earlier comparisons, this one only uses the \citetalias{nml19} SMBH -- host galaxy relation. Because spectroscopic observations are only available for $z < 1$, I use the same \textit{inferred} VDF functions for $1 < z < 1.5$, however, \autoref{fig:Ayr2Dcompare} shows that this redshift range is not a significant contributor to the overall amplitude of the GWB. 

The consistency between the spectroscopic and the \textit{inferred} VDF predictions is promising, implying that the \textit{inferred} VDF is a good proxy for the spectroscopic VDF. As shown in the LEGA-C survey results \citep{LEGAC2022}, the spectroscopic VDF is on the higher end of the error region covered by the \textit{inferred} VDF. Thus, it is not surprising that the spectroscopic VDF predictions are consistent with the higher end of the \textit{inferred} VDF predictions, and the smaller spread is directly attributable to the smaller error bars on the spectroscopic VDF.

\subsection{Prevalence of Individual Sources}

\begin{figure}
    \begin{center}
    \includegraphics[width=\columnwidth]{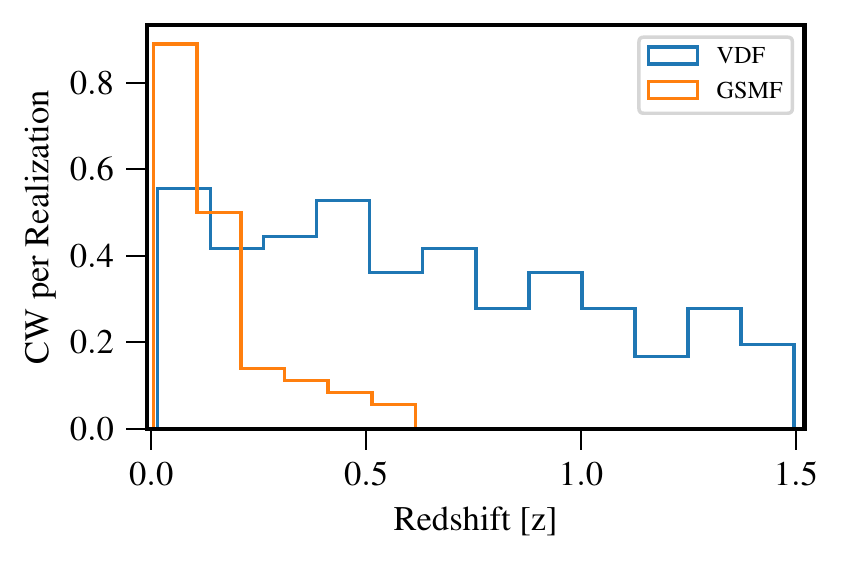}
    \end{center}
    \caption{
    Redshift distribution of individual binary SMBH systems at $f_\mathrm{GW} = 8$ nHz averaged over 100 different population realizations. On average, the VDF method (blue curve) predicts more than twice as many potential individual systems as the GSMF method (orange curve). These systems are expected to be detectable for the IPTA by 2025 \citep{xmh21}. 
    As seen in previous work, the GSMF method predicts that single sources will be found in the local Universe ($z < 0.5$) \citep{rosado+15}. In contrast, the VDF method shows only a slight preference for systems at lower redshift. As seen in \autoref{fig:Ayr2Dcompare}, the larger mass systems at higher redshifts that are present in the VDF predictions directly translate to an increase in individually resolvable sources in PTA data for $z > 0.5$.
    }
    \label{fig:CWsCompare}
\end{figure}

Beyond the amplitude of the GWB, the presence of more massive systems at higher redshifts predicted by the VDF method also has implications for the prevalence of individually resolvable, or continuous wave (CW), sources in PTA datasets. These systems are unique from the GWB, which is compromised of the GW contributions from the entire population of SMBH binary systems in the Universe. CWs are individual systems and once one is detected, it will provide a unique multi-messenger opportunity to study a binary SMBH system. The CWs detectable by PTAs are different than those detectable by other GW experiments, because these systems are typically a long way away from coalescence, and so their frequency evolution should be negligible, allowing PTAs to probe a part of the SMBH binary evolution that is complimentary to other GW experiments. 

\autoref{fig:CWsCompare} shows the redshift distribution of individual binary SMBH systems from one hundred different population realizations using the \citetalias{kh13} SMBH -- host galaxy relations. I use this host galaxy relation since it predicts the most similar amplitudes of the GWB between the two methods, and CW prevalence is directly linked to GWB amplitude \citep{rosado+15}. 
In each case, I have only looked at individual systems that are emitting GWs at a GW frequency close to 8 nHz in order to compare to the expected sky-averaged sensitivity of the IPTA by 2025 \citep{xmh21}. 
On average, the VDF method predicts more than twice as many potential CWs as the GSMF, and the redshift distribution of sources is much broader with the VDF method than with the GSMF. This is to be expected, given the differences in the underlying SMBH populations seen in \autoref{fig:Ayr2Dcompare}. 

The GSMF results shown here are very similar to those in \citet{rosado+15}, which finds that a GWB is more likely to detected before an individually resolvable source and that the first detected individual binary system is likely to be very close by ($z \lesssim 0.5$). 
The VDF results are much more optimistic due to the increase in massive SMBH systems at higher redshift. 
Notably, the VDF method predicts that potentially resolvable sources are fairly evenly distributed across redshift, which is in contrast to other CW predictions, which find that the first detected system is likely to be very near by \citep{rosado+15,kbh+18}. 
The larger systems at higher redshifts have intriguing multi-messenger potential since if they were to have any variable AGN activity they would be promising multi-messenger candidates given near-future time-domain surveys \citep{ctr+21}.

\section{Summary}

In this paper, I undertake a new approach to inferring the SMBH mass function using observations of an \textit{inferred} VDF and compare that to the standard inference method from a GSMF. I also compare the \textit{inferred} VDF to the spectroscopic VDFs that are available within limited redshift ranges.
I am careful to use as many of the same observations as possible between the two methods to ensure that the results are solely due to the different method of inference rather than different observational inputs. 
However, in order to not be biased by the choice of SMBH -- host galaxy relation, which is arguably one of the most impactful parameters \citep[][\citetalias{ss2016}]{sesana13}, I use three separate measurements that span the range of observed values. 
The VDF method predicts a binary SMBH population that is significantly more massive at higher redshifts than the GSMF method. This population is more inline with observed relic galaxies, like NGC1271 and NGC1277 \citep{walsh2015,walsh2016}. 

Interestingly, the VDF method produces $\ayr$ predictions that are at larger amplitudes and cover a narrower spread than the GSMF method, although both methods are able to reproduce the amplitude of the emerging signal in PTA data. The narrower spread in the VDF method's predictions is partially due to the smaller range of $\msig$ measurements, compared to the $\mmb$ relation. This provides further evidence that the $\msig$ relation is a more ``fundamental'' relation between host galaxy's and their SMBHs. 

The largest difference between the two methods is from massive binaries at higher redshifts, where the VDF predicts significant contributions, while the GSMF does not. 
Additionally, the more massive binary SMBHs at higher redshifts predicted by the VDF method increase the number of potential individually resolvable systems in PTA data as well as broaden the redshift range where those systems are expected to be found. This is inline with recent studies of $z \sim 1$ galaxy clusters, which appear to host potential CW systems \citep{wr23}.

It is worth noting that the results presented here are a preliminary investigation into using a VDF to infer the SMBH mass function. The fits to the VDF in \citet{Bezanson+12} are estimates and a full error analysis has yet to be conducted on the \textit{inferred} velocity dispersion functions. However, the consistency between the LEGA-C spectroscopic VDF and \textit{inferred} VDFs is a promising development.
Additionally, the VDFs used in this paper only went out to $z = 1.5$ and as \autoref{fig:Ayr2Dcompare} shows, there is reason to think that significant contributions to the GWB from the VDF method may lie at $z > 1.5$. 
Ideally, the VDF would reach out to $z \sim 2-3$, and work is currently underway to provide it \citep{cayenne23}. 

Finally, I want to reiterate some of the other initial caveats to this work. The model presented here is overly simplistic, looking only at binary SMBHs in circular orbits, ignoring environmental coupling, and assuming a quick, uniform binary evolution timescale. However, these results show that a more direct method of SMBH mass inference is available to binary SMBH populations based on galaxy observations and that the inferred SMBH mass function from this new method is fundamentally different than what has been used in the past. While more work is needed, these results hint towards there being a population of more massive SMBHs at higher redshift than previously thought. 

\begin{acknowledgements}
I am grateful to the referee, Alberto Sesana, for comments and insights that greatly improved this work. 
This project was conceived and advanced through fruitful discussions with Sarah Burke-Spolaor, Julie Comerford, Joseph Lazio, Xavier Siemens, and Sarah Vigeland. The manuscript was improved by comments from Luke Kelley and Kayhan G\"ultekin. I am grateful to Rachel Bezanson for providing key insights to the velocity dispersion function. This work has been supported by the National Science Foundation (NSF) under award 1847938. Additionally, I am supported by an NSF Astronomy and Astrophysics Postdoctoral Fellowship under award AST-2202388.
\end{acknowledgements}

\software{Numpy \citep{numpy},
Scipy \citep{scipy},
MatPlotLib \citep{matplotlib}
}

\bibliography{vdf_bib}
\bibliographystyle{aasjournal}

\end{document}